\documentclass[apjl]{emulateapj}

\usepackage{graphics}

\shorttitle{Two component line profiles of ULX NGC 5408 X-1}
\shortauthors{Cseh et al.}

\begin{document}

\title{Broad Components in Optical Emission Lines from the Ultra-Luminous X-ray Source NGC 5408 X-1}

\author{D. Cseh\altaffilmark{1}, F. Gris\'{e}\altaffilmark{2}, S. Corbel\altaffilmark{1}, P. Kaaret\altaffilmark{2}}
\affil{Laboratoire Astrophysique des Interactions Multi-echelles (UMR 7158), CEA/DSM-CNRS-Universite Paris Diderot, CEA Saclay, F-91191 Gif sur Yvette, France}
\email{david.cseh@cea.fr}

\affil{Department of Physics and Astronomy, University of Iowa, Van Allen Hall, Iowa City, IA 52242, US}

\altaffiltext{1}{Laboratoire Astrophysique des Interactions Multi-echelles (UMR 7158), CEA/DSM-CNRS-Universite Paris Diderot, CEA Saclay, F-91191 Gif sur Yvette, France}
\altaffiltext{2}{Department of Physics and Astronomy, University of Iowa, Van Allen Hall, Iowa City, IA 52242, US}

\begin{abstract}

High-resolution optical spectra of the ultraluminous X-ray source NGC 5408 X-1 show a broad component with a width of $\sim$750~km/s in the He{\sc ii} and H$\beta$ lines in addition to the narrow component observed in these lines and [O {\sc iii}]. Reanalysis of moderate-resolution spectra shows a similar broad component in the He{\sc ii} line. The broad component likely originates in the ULX system itself, probably in the accretion disk. The central wavelength of the broad He{\sc ii} line is shifted by $252 \pm 47$~km/s between the two observations. If this shift represents motion of the compact object,  then its mass is less than $\sim$1800~M$_{\odot}$.
\end{abstract}


\section{Introduction}

Ultraluminous X-ray sources (ULXs) are variable off-nuclear X-ray sources with luminosities exceeding the Eddington luminosity of a $20~M_{\odot}$ compact object, assuming isotropic emission \citep{CM,phil01}. Irregular variability, on time scales from seconds to years, suggests that ULXs contain accreting compact objects. Intermediate mass black holes would be required to produce the inferred luminosities, but ULXs may, instead, accrete at super Eddington rates or be beamed, mechanically or relativistically.

NGC 5408 X-1 is one of the best intermediate mass black hole candidates because it powers a radio nebula requiring an extremely energetic outflow \citep{phil03,soria06,Lang07} and a photoionized nebula requiring an X-ray luminosity above $3\times 10^{39} \rm \, erg \, s^{-1}$ \citep{phil}. Also, quasi-periodic X-ray oscillations at low frequencies suggest a high compact object mass \citep{Strohmayer07}.

The optical counterpart to NGC 5408 X-1 was identified by \citet{Lang07} and optical spectra were obtained by \citet{phil}. The optical spectra had no absorption lines suggesting the emission is not dominated by the companion star.  The observed continuum emission may arise from a nebula or reprocessing of X-rays in an accretion disk.  The optical spectrum is dominated by emission lines, including forbidden lines which must be produced in a low density environment such as a nebula.  Several high excitation lines were detected indicating that the nebula is X-ray photoionized.

\citet{phil} found that the He{\sc ii} line from NGC 5408 X-1 was broader than the forbidden lines. Permitted lines produced in the high-density environment of an accretion disk can be broad, reflecting the distribution of velocities within the optical emitting regions of the disk. Furthermore, since the accretion disk moves with the compact object, the line velocity shifts may provide a means to constrain the compact object mass \citep{Hutchings87,soria98}.

To study the He{\sc ii} line profile of NGC 5408 X-1 in more detail, we obtained new observations using the FORS-2 spectrograph on the European Southern Observatory Very Large Telescope (VLT) with a high resolution grism and reanalyzed our previous FORS-1 observations \citep{phil}.  The observations and data reduction are described in \S 2. The results are presented in \S 3 and discussed in \S 4.

\section{Observations and Analysis}

FORS-2 observations of NGC 5408 X-1 were obtained on 12 April 2010 using the GRIS\_1200B and GRIS\_1200R grisms with a slit width of $1.0\arcsec$ covering the spectral range 3660$-$5110 \AA\ and 5750$-$7310 \AA\ with dispersion 0.36 \AA\ pixel$^{-1}$ and 0.38 \AA\ pixel$^{-1}$ and spectral resolution $\lambda/\Delta\lambda=1420$ and $\lambda/\Delta\lambda=2140$ at the central wavelength, respectively.  The observation block (OB) consisted of three 849~s exposures with a 12 pixel offset along the spatial axis between successive exposures.  CCD pixels were binned for readout by 2 in both the spatial and spectral dimensions. We also reanalyzed all six OBs of the previous FORS-1 observations \citep{phil}, hereafter the low resolution data (LRD), taken using the GRIS\_600B grism which has a spectral resolution of $\lambda/\Delta\lambda=780$ at the central wavelength and with three shifted exposures per OB. The average seeing for our new observations was 0.72 and 0.62 arcsecond for the blue and red spectra, respectively. The average seeing of the six OBs of the LRD were 0.87, 0.82, 0.96, 1.28, 0.64, 0.57 arcsecond, respectively.

Data reduction was carried out using the Image Reduction and Analysis Facility (IRAF)\footnote{IRAF is distributed by the National Optical Astronomy Observatory, which is operated by the Association of Universities for Research in Astronomy, Inc., under cooperative agreement with the National Science Foundation.} \citep{iraf}. First, we created bias and flat-field images, then applied these to correct the spectrum images.  The three exposures in each OB were aligned then averaged to eliminate bad pixels and cosmic rays using the {\tt imcombine} task with the {\tt ccdclip} rejection algorithm.

As the continuum emission of the ULX counterpart is faint, we could not trace its spectrum. Following \citet{phil}, we used the bright nearby star at 2MASS \citep{2mass} position $\alpha_{\rm J2000} = 14^h 03^m 18.^s97, \delta_{\rm J2000} = -41\degr 22\arcmin 56.\arcsec6$ as a reference trace. The trace position on the spatial axis varied less than half a pixel along the whole length of the dispersion axis. The trace for the ULX counterpart was centered on the He{\sc ii} $\lambda 4686$ emission line profile. The smallest possible trace width, 2 pixels corresponding to $0.5\arcsec$, was used to best isolate the ULX emission from the nebular emission.  Background subtraction was done with a trace close by. The HgCdHeNeA lamp and standard star LTT7379 were used for wavelength and flux calibration. An atmospheric extinction correction was applied using the IRAF built-in Cerro Tololo Inter-American Observatory (CTIO) extinction tables.  To estimate the reddening, we used the Balmer decrement of H$\delta$/H$\beta$, we find $E(B-V)=0.08 \pm 0.03$ in agreement with \citet{phil}. We corrected for reddening using the extinction curve from \citet{dered} with $R_{V}=3.1$.

To study the kinematics, we need to characterize the instrumental resolution in order to obtain intrinsic line widths. After applying the dispersion correction to the lamp spectrum, we measured the full width at half maximum (FWHM) of several lines, excluding saturated ones, by fitting Gaussians with the IRAF {\tt splot} subroutine. The instrumental FWHM was 2.24 \AA \,and 5.08 \AA \,for the high and low resolution data, respectively.  The error on the instrumental FHWM was estimated by finding the standard deviation of the FWHM for several different lines.

For the He{\sc ii}, H$\beta$ and [O{\sc iii}] emission lines, see Fig.~1. and Table~\ref{fit}, we first fitted the continuum with a second order polynomial to a region around each line excluding the line itself by visual examination. We estimated the measurement errors by calculating the root mean square deviation of the data in the same region. Then, we performed a non-linear least squares fit using the {\tt LMFIT} subroutine of the Interactive Data Language version 7.0  and based on "MRQMIN" \citep{numrec}. We fitted the line profiles iteratively, first using one Gaussian which converged on the narrow component, then using a sum of two Gaussians with initial parameters adjusted to achieve convergence. All six parameters in the two Gaussian fit were  free to vary.  The errors on the parameters were calculated by the fitting routine  in a way that the uncertainty for the $i$th parameter derives from the
square-root of the corresponding diagonal element of the covariance matrix.  The intrinsic line width was calculated assuming the measured line width is the quadrature sum of the intrinsic and instrumental widths and the error on the intrinsic line width included a term for the uncertainty in the instrumental FHWM.

For the H$\alpha$ line, we fitted the sum of four Gaussians, because the [N{\sc ii}] lines lie on the red and blue parts of the line wing. Initial fits to the H$\alpha$ and red [N{\sc ii}] lines provided initial values for a fit with four Gaussians. Because the blue [N{\sc ii}] line has very low signal to noise ratio, the widths of the two [N{\sc ii}] lines were set equal, the wavelength offset was fixed at $-35.44$~\AA, and the amplitude of the blue line was set to 1/3 of the red line \citep{oster}.

\begin{figure}\label{new}
\center
%\rotatebox{90}{
\includegraphics[width=3.5in]{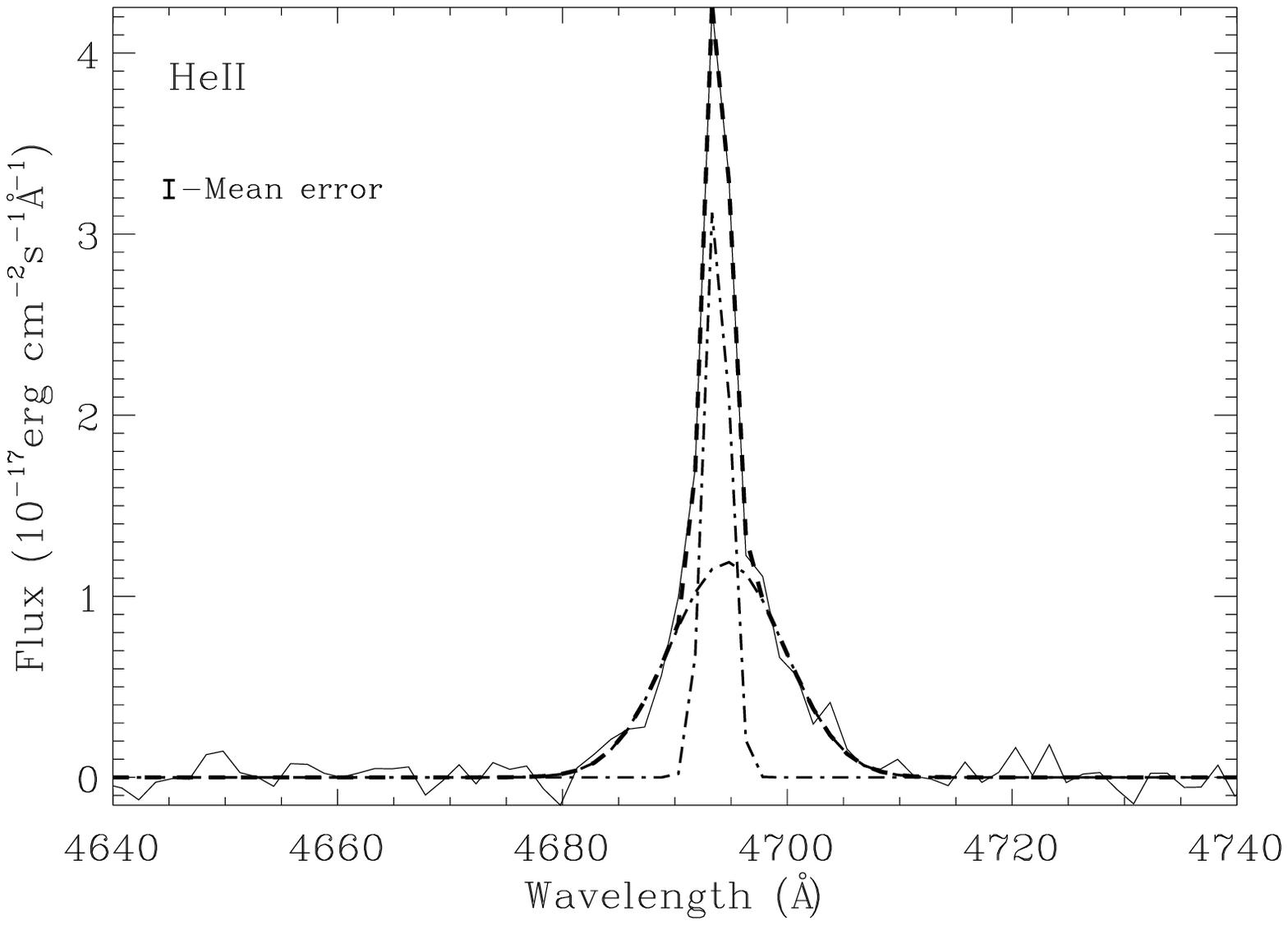}
\includegraphics[width=3.5in]{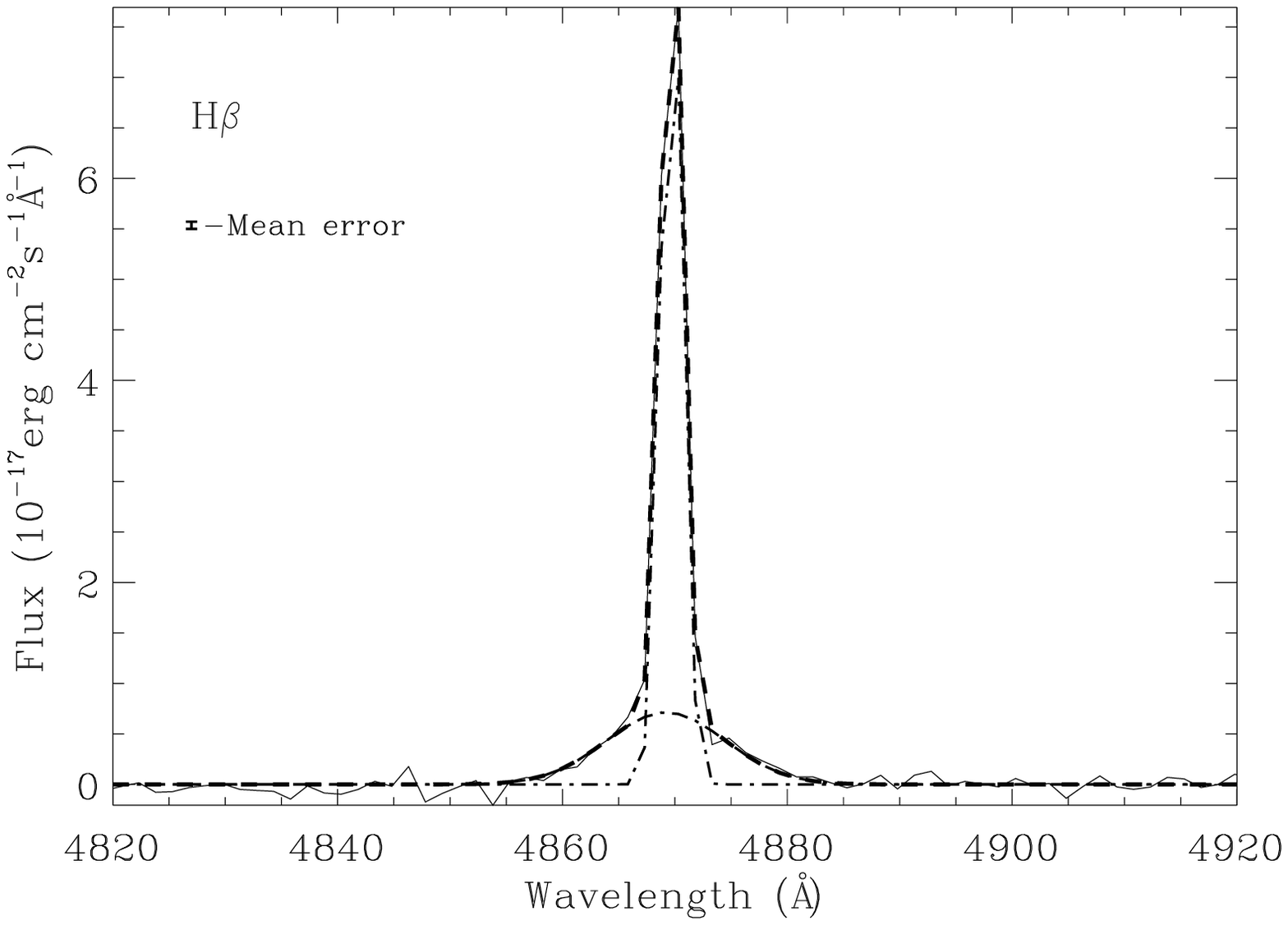}
\includegraphics[width=3.35in]{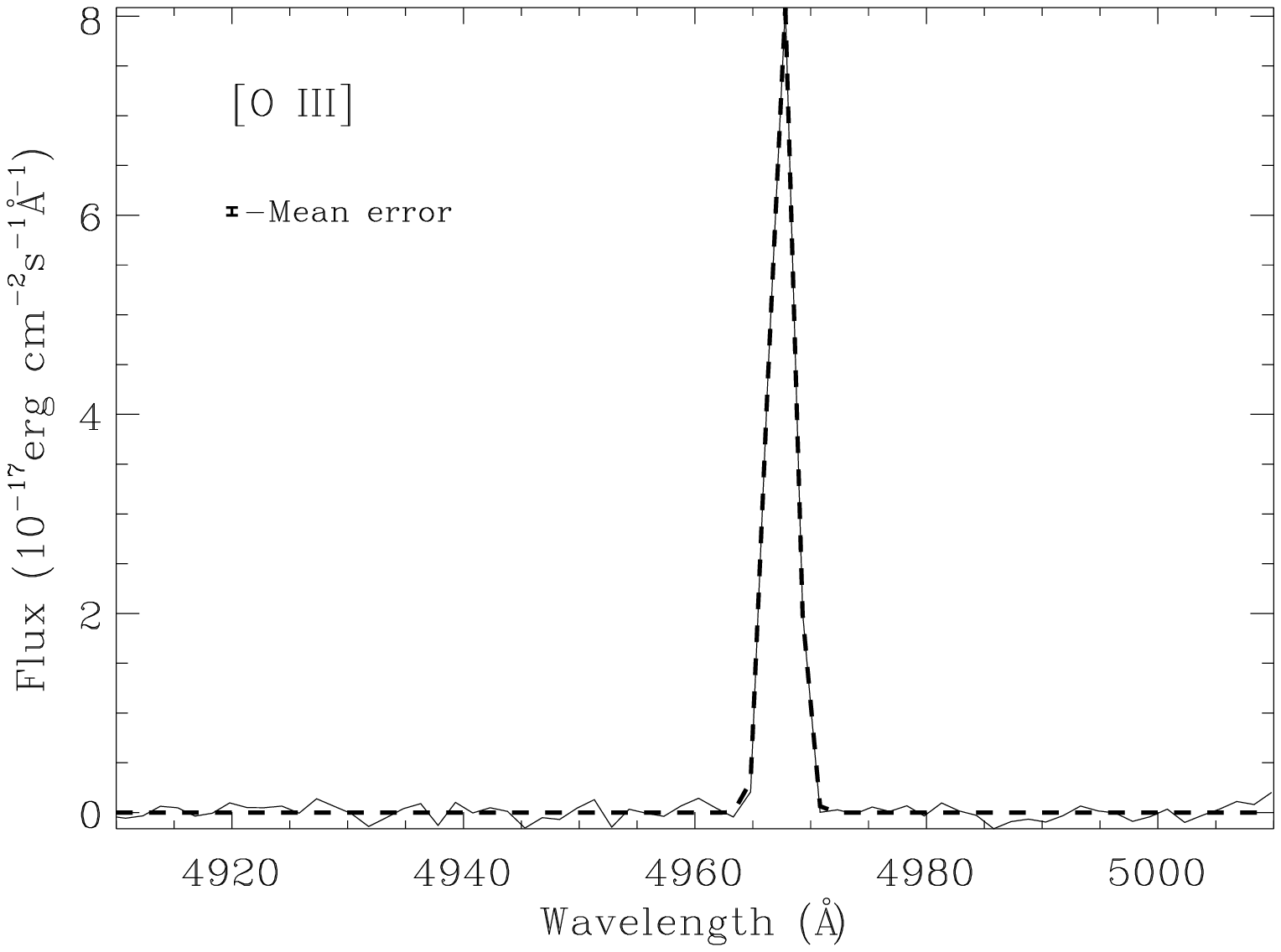}%}
\caption{He{\sc ii}, H$\beta$ and [O {\sc iii}] lines from the new, high resolution data. Results of line fits with two Gaussian components are shown as the separate components plotted with dashed-dot lines and the overall fit plotted with dashed lines (except [O {\sc iii}]).}
 \end{figure}

\begin{figure}\label{LRD}
\center
%\rotatebox{90}{
\includegraphics[width=3.5in]{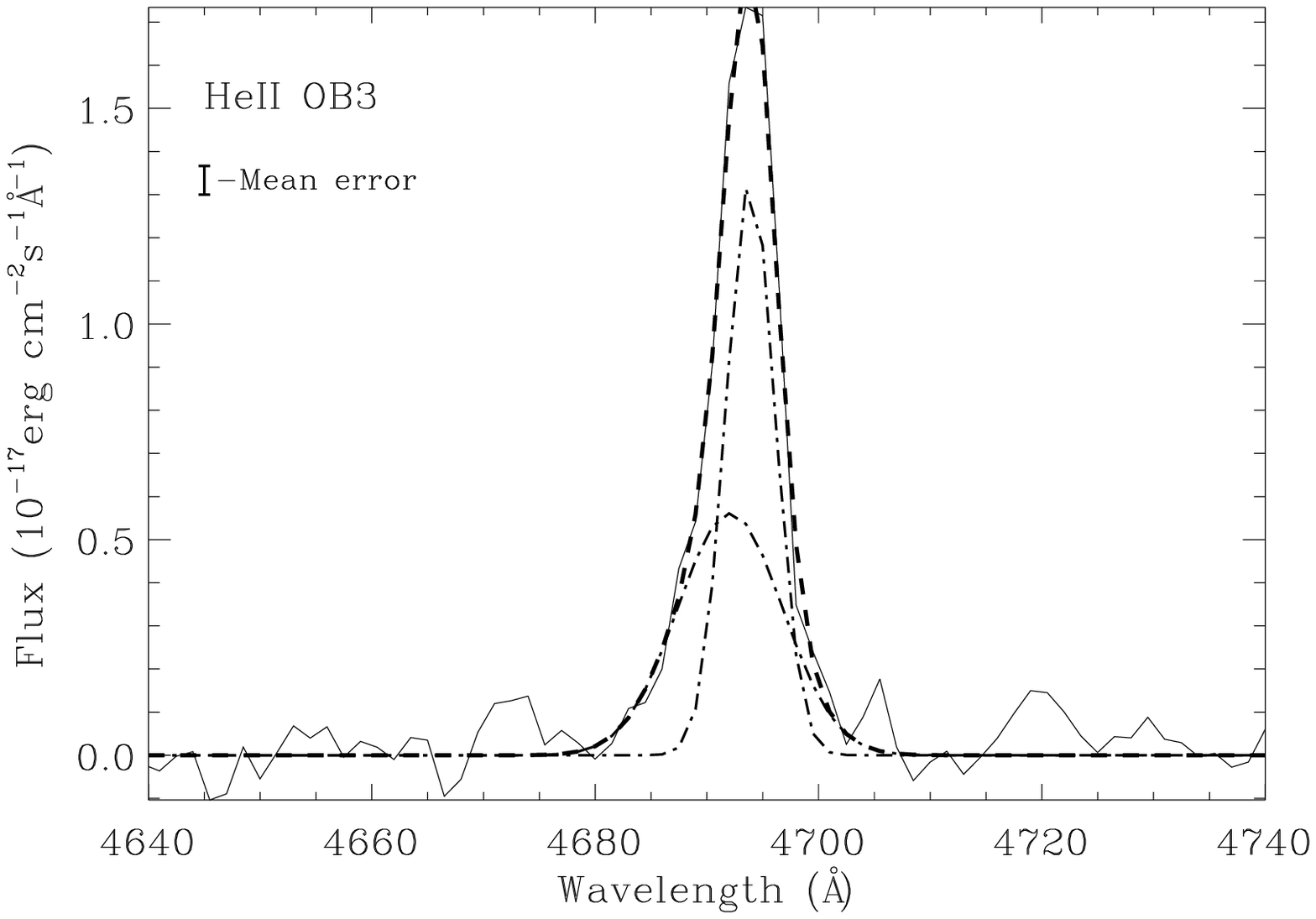}
\includegraphics[width=3.5in]{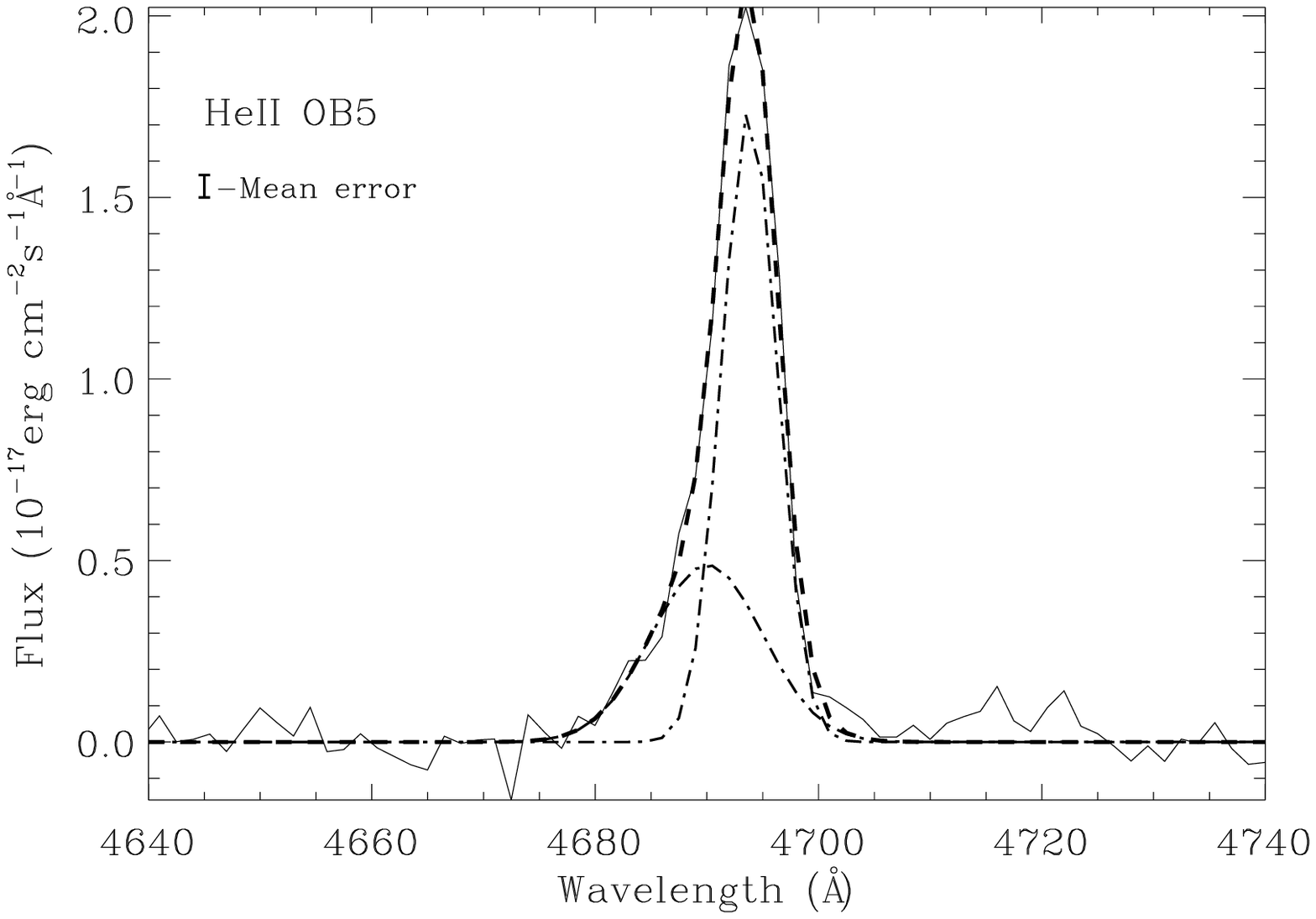}
\includegraphics[width=3.5in]{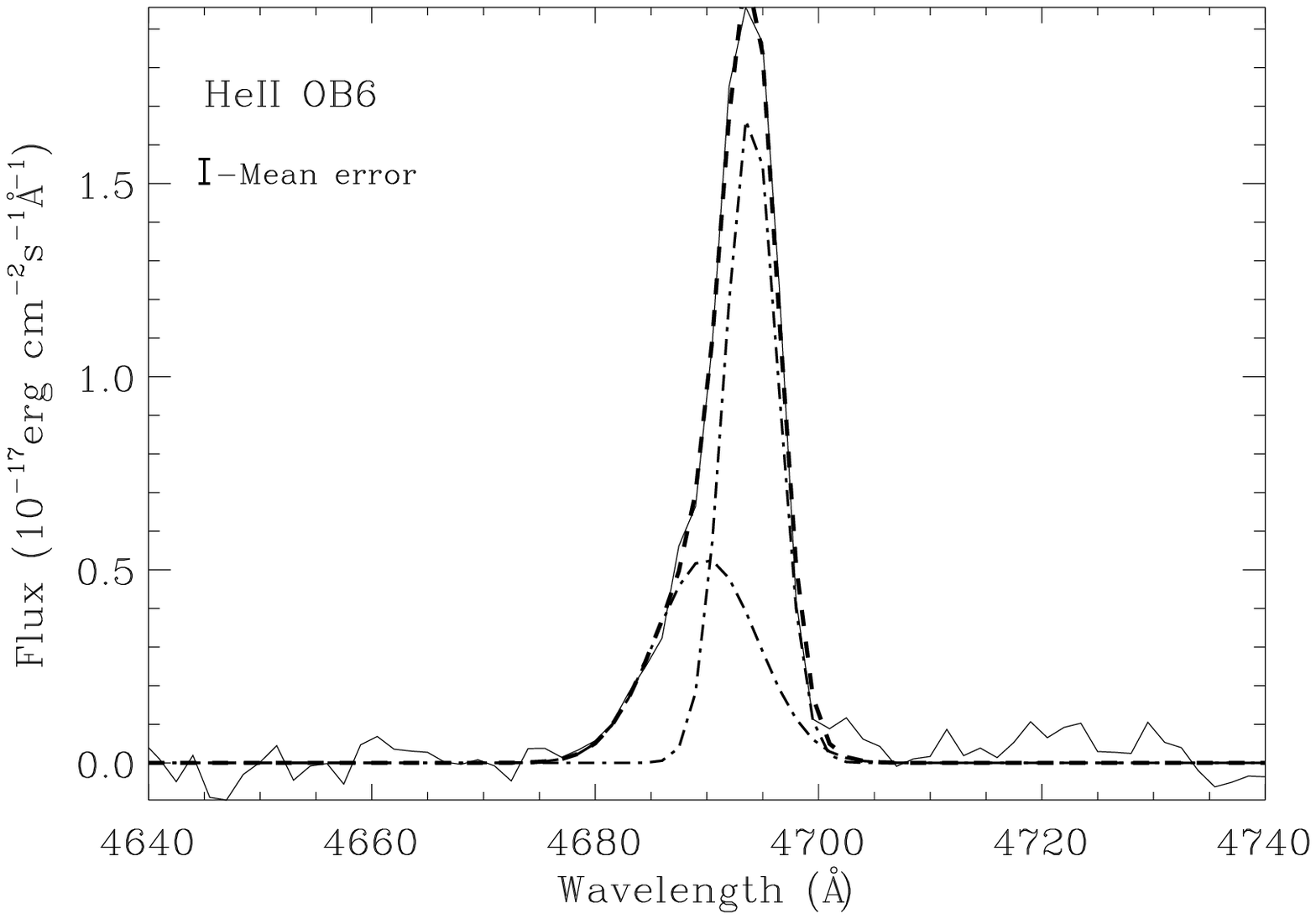}%}
\caption{He{\sc ii} line from the low resolution data(OB3,OB5,OB6). Results of line fits with two Gaussian components are shown as the separate components plotted with dashed-dot lines and the overall fit plotted with dashed lines.}
 \end{figure}

\section{Results}

The improved resolution of the new data clearly resolves a broad component in the He{\sc ii} line profile, see Fig.~1. and Table~\ref{fit}. The centroid is shifted from the nebular component by $+0.87\pm0.26 \rm \, \AA$ in the red direction. We also searched for broad components in other lines. H$\beta$ has a broad component with a FWHM similar to the He{\sc ii} line but shifted by $-0.52\pm0.32 \rm \, \AA$ towards the blue, rather than the red. In contrast, a single Gaussian provides a good fit to the forbidden [O{\sc iii}] line and there is no evidence for a broad component, as expected if the line is emitted only from the nebula.

Then we fitted the He{\sc ii} line profiles of the six OBs of the LRD, see Fig.~2. and Table~\ref{fit}. The flux variation of the overall line profiles correlates with the seeing, e.g.\ OB5 has the best seeing and the highest flux. We detected a broad component in the He{\sc ii} line in OB3, OB5, and OB6. We did not significantly detect a broad component in OB1, OB2 and OB4.  This may be due to seeing or variations in the flux of the broad component.

\begin{table*}
\begin{center}
\caption{Line fit results.\label{fit}}
\begin{tabular}{lccccc}
\tableline\tableline

Line & Wavelength (\AA)  & FWHM (\AA) & Flux  & Velocity (km s$^{-1}$) & $\chi^2_{\nu}$\\
\tableline
He{\sc ii}	&	4693.75	$\pm$	0.04	&	2.54	$\pm$	0.14	&	3.39	$\pm$	0.13	&	76.3	$\pm$	3.4	&	0.94\\
	&	4694.62	$\pm$	0.26	&	11.98	$\pm$	0.74	&	1.19	$\pm$	0.10	&	752.0	$\pm$	46.4	&	\\
H$\beta$	&	4869.73	$\pm$	0.01	&	2.28	$\pm$	0.05	&	8.37	$\pm$	0.11	&	25.2	$\pm$	3.3	&	1.08\\
	&	4869.21	$\pm$	0.32	&	12.49	$\pm$	0.94	&	0.71	$\pm$	0.07	&	756.9	$\pm$	57.5	&	\\
$[$O \sc{iii}] $\lambda$4959	&	4967.52	$\pm$	0.01	&	2.48	$\pm$	0.03	&	8.42	$\pm$	0.08	&	64.6	$\pm$	1.7	&	1.08\\
\tableline
He{\sc ii} OB3	&	4693.90	$\pm$	0.15	&	5.15	$\pm$	0.52	&	1.34	$\pm$	0.19	&	52.0	$\pm$	33.3	&	1.04\\
	&	4692.08	$\pm$	0.77	&	11.09	$\pm$	1.44	&	0.56	$\pm$	0.20	&	630.5	$\pm$	91.8	&	\\
He{\sc ii} OB5	&	4693.80	$\pm$	0.12	&	5.81	$\pm$	0.41	&	1.74	$\pm$	0.17	&	179.8	$\pm$	25.7	&	1.03\\
	&	4690.03	$\pm$	1.43	&	11.75	$\pm$	1.33	&	0.49	$\pm$	0.13	&	678.0	$\pm$	84.7	&	\\
He{\sc ii} OB6	&	4693.97	$\pm$	0.12	&	5.56	$\pm$	0.46	&	1.70	$\pm$	0.23	&	143.7	$\pm$	29.5	&	0.95\\
	&	4689.94	$\pm$	1.70	&	10.82	$\pm$	1.55	&	0.53	$\pm$	0.15	&	610.7	$\pm$	98.9	&	\\
He{\sc ii} AVG	&	4693.89	$\pm$	0.08	&	5.50	$\pm$	0.26	&	1.58	$\pm$	0.10	&	135.0	$\pm$	16.6	&	1.15\\
	&	4690.82	$\pm$	0.68	&	11.47	$\pm$	0.73	&	0.53	$\pm$	0.09	&	657.5	$\pm$	46.3	&	\\
\end{tabular}
\tablecomments{The table gives the: line identification, central wavelength, measured FWHM, flux in units of  10$^{-17}$ erg cm$^{-2}$ s$^{-1}$, intrinsic FWHM in velocity units, and $\chi^2_{\nu}$ of the line profile fit. The first row for each spectral line shows the parameters of the narrow component and the second row shows the parameters of the broad component (except for [O {\sc iii}]).} 
\end{center}
\end{table*}

We note that \citet{phil} reported lower fluxes for He{\sc ii}, [Ne{\sc v}], and the continuum emission for OB4 (with by far the worst seeing) as compared to the other OBs, while the other line fluxes remained relatively constant. Our new analysis suggests that this is due to changes in the seeing. If the emitting region is smaller than the $0.51\arcsec$ slit used for the LRD, then poor seeing will decrease the flux through the spectrometer.  If these emission components are enhanced close to the ULX system while the other line emission is uniform, then the poor seeing in OB4 would produce the observed changes in flux. Thus, there is no evidence for temporal variability of the continuum or line emission.  However, the subtraction performed by \citet{phil} to isolate continuum emission arising from near the ULX is still justified; the separation of components is spatial instead of temporal.

The He{\sc ii} line parameters are consistent between OB3, OB5, and OB6. The wavelength shifts of the He{\sc ii} broad component of OB3, OB5 and OB6 relative to the narrow component are $-1.82\pm0.78$ \AA, $-3.77\pm1.44$ \AA$\,$ and $-4.03\pm1.70$ \AA \,into the blue direction instead of the red as in the HRD, and are consistent within one $\sigma$.  We averaged the spectra for these three observations and fit the resulting line profile.  The fit results are listed as He{\sc ii} AVG in Table~\ref{fit}.  The shift of average line profile is $-3.07\pm0.68$ \AA.

The He{\sc ii} broad component width is consistent between the new and old data.  The narrow component is wider in the old data because we do not resolve the nebular lines. The line fluxes are higher in the new data, most likely due to the wider slit.  The centroids of the narrow component are consistent, while the wavelength shift of the He{\sc ii} broad component between the old and new data is $\Delta\lambda=3.94\pm0.73$~\AA.

 Fitting the H$\beta$ line of the LRD, we did not get a good fit, due to the lack of spectral resolution and the low broad to narrow flux ratio.  We could not fit the bluer Balmer lines because of their low S/N ratios.  We did fit the H$\alpha$ line in the new data.  Although we do not obtain a good fit ($\chi^2_{\nu}=4.9$) because of the complicated line profile (i.e.\ the two [N{\sc ii}] lines lie on the red and blue wing of the H$\alpha$ line), we find that there is a broad component with a width of 19 \AA, while the width of the nebular component is 2.7~\AA.  The [N{\sc ii}] lines are narrow, with a typical width of 3 \AA, quantitatively supporting that the forbidden, nebular lines do not have broad components.

\section{Discussion}

Our new, high-resolution spectra show narrow nebular lines and broad components in the He{\sc ii}, H$\beta$, and H$\alpha$ lines.  Our previous, moderate-resolution spectra show a broad component in the He{\sc ii} line. There is no broad component in the [O {\sc iii}] nebular lines in either the new or old spectra. There is still no sign of any absorption lines in the new spectra.

\subsection{The line emitting region}

The broad components of both He{\sc ii} and H$\beta$ have widths $\sim$750 km/s, consistent with production in the accretion disk, and are roughly Gaussian, instead of having P-Cygni profiles that would indicate origin in a wind.  Following \citet{porter}, we estimate the size of the line-emitting region, $R_{le}$, by assuming the line-emitting gas is in Keplerian orbits around a compact object, thus $R_{le} \le GM/v^2$.  We find $R_{le}<2.35 \left( \frac{M_{\rm{BH}}}{1500\, \rm{M}_{\odot}} \right)$~AU, which for a mass of 10 M$_{\odot}$ would give an upper limit of 3.4 R$_{\odot}$. This is consistent with origin of the broad He{\sc ii} line in the accretion disk.

The broad line components are shifted relative to the narrow components.  In the new data, the shifts are small compared to the line width, $+56 \pm 17$~km/s for He{\sc ii} and $-33 \pm 20$~km/s for H$\beta$.  These shifts are consistent only at the $3\sigma$ level, which might indicate a difference in the spatial origin of the lines.  However, this is still consistent with production of both lines within the disk since random motions within the disk and variation between the emission regions could produce shifts that are small compared to the line widths, as observed.

The central wavelength of the He{\sc ii} broad component shifts markedly between the odld and new data, $\Delta\lambda=3.94\pm0.73$ \AA$\,$ or $\Delta v = 252 \pm 47 $~km/s.  This shift is a substantial fraction of the line width. The shift could be due to random motion within the disk, differing viewing geometries \citep{rob10}, or orbital motion of disk (and the compact object).  If the shifts in the broad component of the He{\sc ii} line are due to orbital motion, then this would provide a means to determine the orbital period and would also provide a measurement of the mass function for the secondary star.  Thus, a program of monitoring NGC 5408 X-1 with high-resolution optical spectroscopic observations will be important in extending our understanding of the physical nature of this system.

\subsection{The binary system}

In this section, we make some speculations based on interpretation of the shift in the broad component of the He{\sc ii} line as due to orbital motion. One can express the mass function and the compact object mass, $M_x$, in terms of the orbital period, $P$, the velocity excursion, $K_{x}$, and the companion mass, $M_{c}$, as

\begin{equation}
f_x = \frac{PK_{x}^3}{2\pi G}=\frac{M_{c} \sin^{3}i}{(1+\frac{M_{x}}{M_{c}})^2} \le M_c
\end{equation}

\begin{figure}\label{binary}
\center
\includegraphics[width=3.5in]{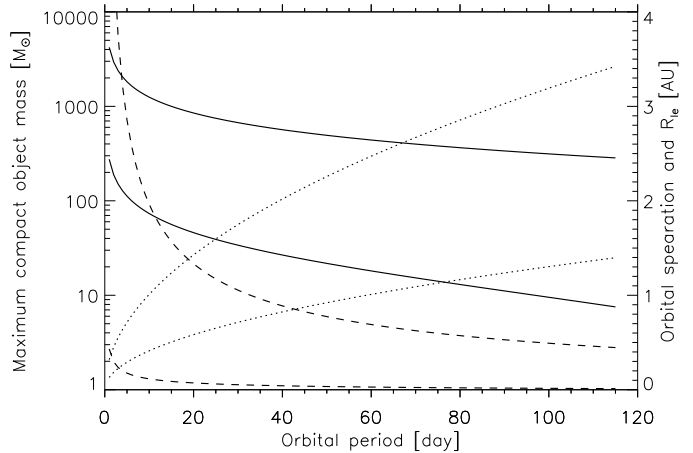}
\caption{Properties of the binary system as a function of orbital period.  The two solid lines show the mass of the compact object; the upper line is for an 120~M$_{\odot}$ donor and the lower line is for a 20~M$_{\odot}$ donor (left vertical scale). The orbital separation is shown as dotted lines with values read on the right vertical scale; the upper line for an 120~M$_{\odot}$ and the lower is for 20~M$_{\odot}$ donor (right vertical scale).  The two dashed lines show the upper limit on the size of the line-emitting region; the upper line is for an 120~M$_{\odot}$ donor and the lower line is for a 20~M$_{\odot}$ donor (right vertical scale). The intercept of the dotted and dashed lines provides a limit on the period and consequently on the mass of the black hole.}
 \end{figure}

\begin{equation}
M_x = \left(M_c \sin i\right)^{\frac{3}{2}} \left(\frac{PK_{x}^3}{2\pi G}\right)^{-\frac{1}{2}} - M_c
\end{equation}

\noindent  where $i$ is the inclination angle and $G$ is the gravitational constant. From the shift of the He{\sc ii} line quoted above, we constrain the semi-amplitude of the radial velocity $K_x \ge \Delta v/2 = 126 \pm 24$~km/s.  Thus, if the maximum mass of the companion and the orbital period are known, then Eq.~2 leads to an upper bound on the mass of the compact object.

 The binary system has a visual magnitude $v_0=22.2$ that gives an upper limit on the absolute magnitude of the companion of $V_0=-6.2$ at a distance of 4.8 Mpc \citep{dist}. Unfortunately, this places little restriction on the companion mass as even O3V stars, with masses of 120~$M_{\odot}$, are allowed.  However, very high mass stars are very short lived, no more than a few million years.  There is no evidence of a dense stellar association near NGC 5408 X-1 and origin in the closest super-star cluster would require a transit time to the present location on the order of 30~Myr \citep{phil03}.  Thus, the companion mass is likely significantly lower, near 20~$M_{\odot}$ or less similar to found from studies of the stellar environments of other ULXs \citep{fabien08,fabien11}.  Figure~3. shows the upper bound on the compact object mass for donors of 120~M$_{\odot}$ and 20~M$_{\odot}$ as a function of orbital period. High black hole masses are excluded, except for very short periods. We note that the He{\sc ii} line shift was the same in OB3 versus OB5 and OB6, taken one day apart, suggesting that the period is longer than few days. Thus, the black hole mass is likely below $\sim$1800~M$_{\odot}$.  The more probable companion mass of 20~M$_{\odot}$ or less would imply smaller black hole masses, less than 112~M$_{\odot}$.

As a further constraint, we note that the orbital separation should be larger than the size of the emitting region, calculated above.  Assuming a circular orbit, the orbital separation of the compact object is $a = \left(\frac{G(M_c +M_{x})P^2}{4\pi^2} \right)^{\frac{1}{3}}$.  Figure~3. shows the the orbital separation as a function of period. Also shown is the size of the line-emitting region, $R_{le}$, versus period. Both are calculated using the maximum black hole mass for each period assuming an 120~$M_{\odot}$ or 20~$M_{\odot}$ donor. The orbital separation is greater than the upper limit of the size of the line-emitting region when the compact object mass is below 875~M$_{\odot}$ for an 120~$M_{\odot}$ companion and below 128~$M_{\odot}$ for a 20~$M_{\odot}$ companion. These masses are reduced if the inclination is lowered. These results suggest that the most probable black hole mass is at most a factor of several above the usual stellar-mass black hole range.

 \citet{tod} proposed an orbital period of  $P=115.5\pm4.0$ days for NGC 5408 X-1, based on variations in the X-ray emission.  With $P=115.5$ days, the mass function is $f_x=24.0 \pm$  13.4~M$_{\odot}$, implying a lower bound on the companion mass $M_c \ge$  10.6~M$_{\odot}$.  It is interesting to determine if this period is consistent with other constraints on the system.  An orbital period of  $P=115.5\pm4.0$ days would require a mean stellar density of $\rho=1.5\times10^{-5}$ g cm$^{-3}$ and, thus, a supergiant companion if mass transfer proceeds via Roche-lobe overflow \citep{tod,phil06}.  In particular, late F and early G supergiants have densities close to that required, although we caution that the high mass transfer rate needed to power the ULX may distort the spectral type of the star.  Such stars have masses of 10--12$M_{\odot}$, consistent with the minimum mass derived from the mass function, and absolute magnitudes close to or below the upper limit quoted above. The stellar radii are large, up to 0.7~AU, but smaller than the orbital separation for this period and mass. However, a companion mass so close to the lower bound on the mass function would require a very low mass compact object.  For a companion mass of 10--12~$M_{\odot}$, the compact object would have to be below 1~$M_{\odot}$, which seems unlikely.  For a 5~$M_{\odot}$ black hole, one would need a donor of about 17~$M_{\odot}$ if the system is edge on.  A higher black hole mass or a less extreme inclination would require an even higher mass companion.  These high companion masses contradict the required stellar density; there is no star with both $M_{c} \ge 17~M_{\odot}$ and $\rho \sim 1.5 \times 10^{-5}$ g cm$^{-3}$.  Thus, either the orbital period is not near 115~days or mass transfer does not proceed via Roche-lobe overflow.  We note that \citet{superorb} have suggested that 115~day periodicity may, instead, indicate a super-orbital period.

\acknowledgments

We thank Manfred Pakull and the referee for valuable comments. This research received funding from the European Community's Seventh Framework Programme (FP7/2007-2013) under grant agreement number ITN 215212 "Black Hole Universe". We acknowledge financial support from a Marie Curie Reintegration Grant under contract number PERG05-GA-2009-243869 and from NASA grant NNX08AJ26G.{\it Facilities:} \facility{ESO VLT}.

\end{document}